\documentclass[a4paper,fleqn,usenatbib]{mnras}
\usepackage{subfigure}
\usepackage{rotating}
\usepackage{newtxtext,newtxmath}
\usepackage[T1]{fontenc}
\usepackage{ae,aecompl} 
\usepackage{tikz}
\usepackage{amsmath}
\usepackage{nccmath}
\usepackage{multicol, blindtext}

\usepackage{graphicx}	
\usepackage{amsmath}	
\usepackage{float}		
\usepackage{placeins}
\usepackage{array}
\usepackage{tabularx}

\newcommand{\Ms}{M$_{\odot}$}

\title[The role of previous generations of stars]{The role of previous generations of stars in triggering star formation and driving gas dynamics}
\author[Herrington et al.]{Nicholas P. Herrington,$^{1}$\thanks{E-mail: nicholas.herrington.astro@gmail.com} 
Clare L. Dobbs,$^1$
Thomas J. R. Bending$^{1}$
\\
\\
$^1$School of Physics and Astronomy, University of Exeter, Stocker Road, Exeter, EX4 4QL, UK \\}

\date{Accepted XXX. Received YYY; in original form ZZZ}

\begin{document}
\label{firstpage}
\pagerange{\pageref{firstpage}--\pageref{lastpage}}
\maketitle

\begin{abstract}
We present hydrodynamic and magnetohydrodynamic (MHD) simulations of sub galactic regions including photoionising and supernova feedack. We aim to improve the initial conditions of our region extraction models by including an initial population of stars. We also  investigate the reliability of extracting regions in simulations, and show that with a good choice of region, results are comparable with using a larger region for the duration of our simulations.
Simulations of star formation on molecular cloud scales typically start with a turbulent cloud of gas, from which stars form and then undergo feedback.
 In reality, a typical cloud or region within a galaxy may already include, or reside near some population of stars containing massive stars undergoing feedback. 
 We find the main role of a prior population is triggering star formation, and contributing to gas dynamics. 
 Early time supernova from the initial population are important in triggering new star formation and driving gas motions on larger scales above 100 pc, whilst the ionising feedback contribution from the initial population has less impact, since many members of the initial population have cleared out gas around them in the prior model. In terms of overall star formation rates though, the initial population has a relatively small effect, and the feedback does not for example suppress subsequent star formation. We find that MHD has a relatively larger impact than initial conditions, reducing the star formation rate by a factor of 3 at later times.

\end{abstract}


\section{Introduction}
Most if not all star formation is influenced by larger scale processes such as spiral arms, bars, feedback from massive stars and large scale turbulence. Spiral arms induce converging flows \citep{Roberts1969}, bars drive gas flows towards the centres of galaxies \citep{Krumholz2017,Seo2019} and are associated with large shearing motions \citep{Sheth2000,Khopersov2018}, whilst star formation is often supposed to be triggered within dense shells from HII regions and supernovae (e.g. \citealt{Elmegreen1977,Tieftrunk1997,Elmegreen1998,Thompson2004,Martins2010,Smith2010,Deharveng2012}). Star formation may thus be influenced by larger scale dynamics, as well as the past star formation history of a region or a galaxy, whereby previous generations of stars may be driving turbulence, or producing shells and bubbles in the ISM.

Incorporating these processes in numerical studies of star formation is challenging. Many simulations do not attempt to explicitly include larger scale processes, and simply model isolated clouds subject to turbulence driven on larger scales \citep{Bate2002,Clark2006,Bate2009,Girichidis2011,Federrath2014}, or converging flows which are assumed to be from spiral arms, supernovae feedback, or cloud collisions \citep{heitsch2006,hennebelle2008,Banerjee2009,ntormousi2011,Duarte-Cabral2011,clark2012,Walch2012,balfour2015,dobbs2015cc,fogerty2016,dobbs2020,liow2020}. These not only miss the larger scale dynamics, but also that regions of star formation may occur near to young massive stars, with such stars undergoing ionisation, winds and supernovae on different timescales. Observationally, we see evidence for sequential star formation \citep{Fukuda2000,Deharveng2003,Preibisch2007}, suggesting that later groups of clusters of stars will be forming in or close to regions subject to feedback and where supernovae have already occurred.

An alternative approach is to simulate regions from larger scale simulations at higher resolution \citep{rey-raposo2015,dobbs2015re,shima2017,seifried2017,rey-raposo2017,Bending2020,Ali2021,rieder2020,bending2022,rieder2022,dobbs2022}. These simulations have the advantage that they capture the resolution necessary to follow low mass molecular clouds, star formation within molecular clouds, feedback processes and their effect on star formation, as well as spiral arm dynamics, or galaxy collisions. Recent work on these scales is able to follow cluster evolution, showing that interactions with the surroundings, e.g. incoming gas and other clusters, as well as hierarchical cluster formation whereby multiple clusters are forming, and often merging together, are important for cluster formation and evolution, particularly for more massive clusters and longer term cluster evolution \citep{rieder2022,dobbs2022}. A common problem with these previous works, both the isolated clouds and galactic region scale simulations however are the boundary conditions, which are usually assumed to be empty, or for grid codes, periodic, and again the prior star formation history (and particularly whether there are any nearby massive stars) is typically neglected. It is not clear what difference missing this information makes to the validity of these models.

Other works, particularly with moving mesh codes, have instead performed zoom-in simulations where the rest of the galaxy is modelled at lower resolution, and one region is modelled at much higher resolution \citep{Smith2020,Li2022}. However this technique may be less suitable for smoothed particle hydrodynamics (SPH) since this would involve adjacent particles, or mixing of particles with very different masses. For all codes, these simulations also may be unnecessarily computationally expensive, or memory intensive, since the whole galaxy is also modelled but not of interest. And there is still a difference in resolution that cannot be avoided, so for example feedback which occurs outside the high resolution region will only be modelled with low resolution.

The star formation rate is one specific property that may depend on the boundary conditions and prior stellar population.
In a recent paper, \citet{Bending2020}, we found the surprising result that photoionising feedback appeared to slightly increase rather than decrease star formation. One important difference in this simulation compared to previous work was that the simulation was somewhat larger scale, so rather than feedback simply propagating into a vacuum, feedback causes new star formation to occur. We also saw that star formation initially increased then decayed compared to having no feedback, and postulated that this behaviour may be due to, or exacerbated by not having any initial population of stars. Rather than feedback acting initially to reduce star formation, stars form and then produce feedback. We supposed that including an initial population of stars might affect the ISM at earlier times before the first burst of star formation occurs, possibly regulating star formation better. As well as feedback, we also expect that magnetic fields, which were not included in \citet{Bending2020} will influence the amount of star formation which occurs in the simulations.

In this paper we examine how sensitive results from high resolution simulations of isolated galaxy regions are to the inclusion of the surrounding interstellar medium, i.e. testing the reliability of extracting a smaller region from a large scale simulation, and whether including prior populations of stars is something we need to do when modelling star formation. We also look at whether feedback from an initial population triggers or reduces star formation. Related to this we also compare how the inclusion of feedback generally, the inclusion of a prior stellar population and the inclusion of magnetic fields affect the star formation rates and efficiency of star formation in our regions.

\section{Methodology}

\begin{figure} 
\begin{tabular}{c}
\includegraphics[width=0.46\textwidth]{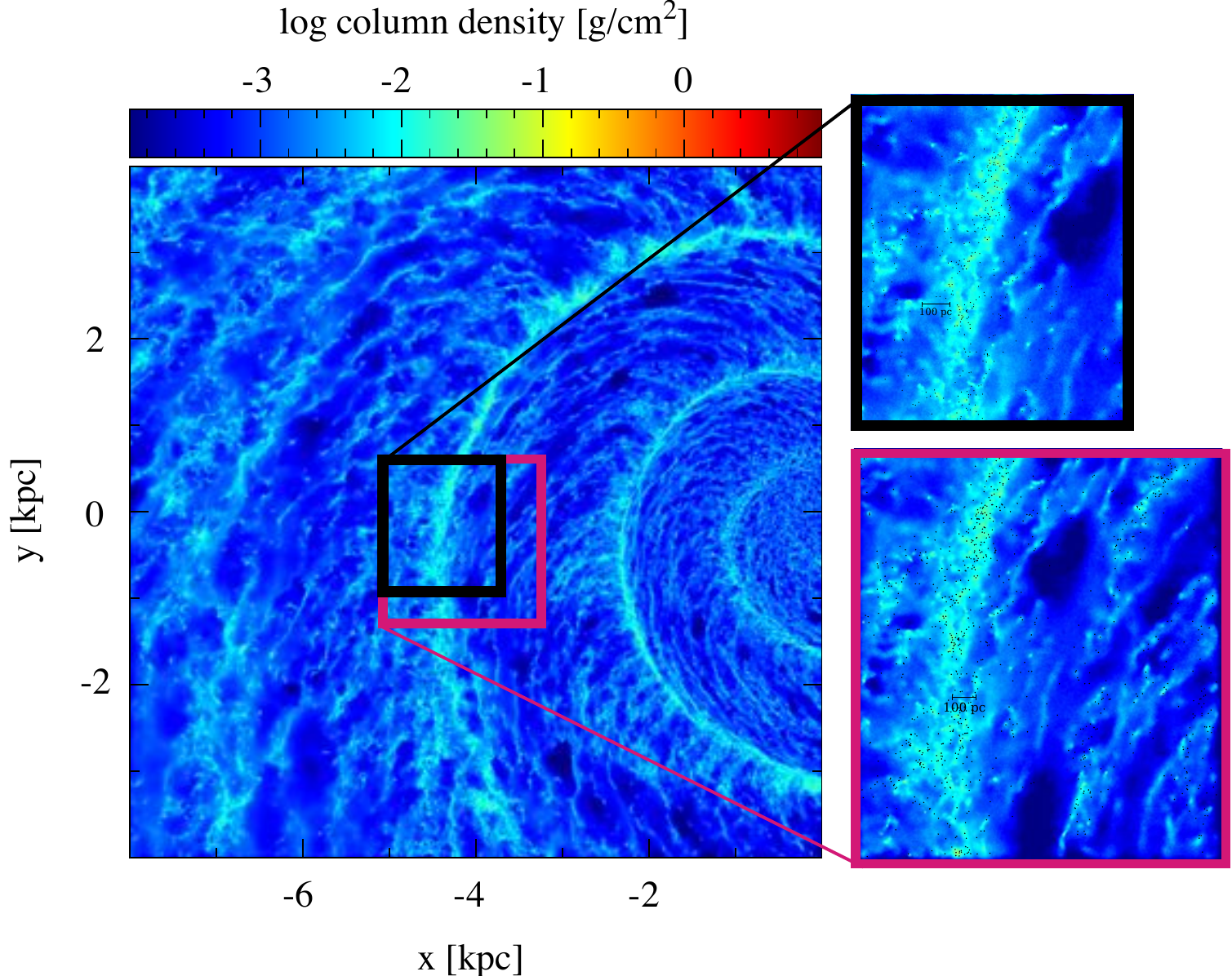} \\ 
\includegraphics[width=0.46\textwidth]{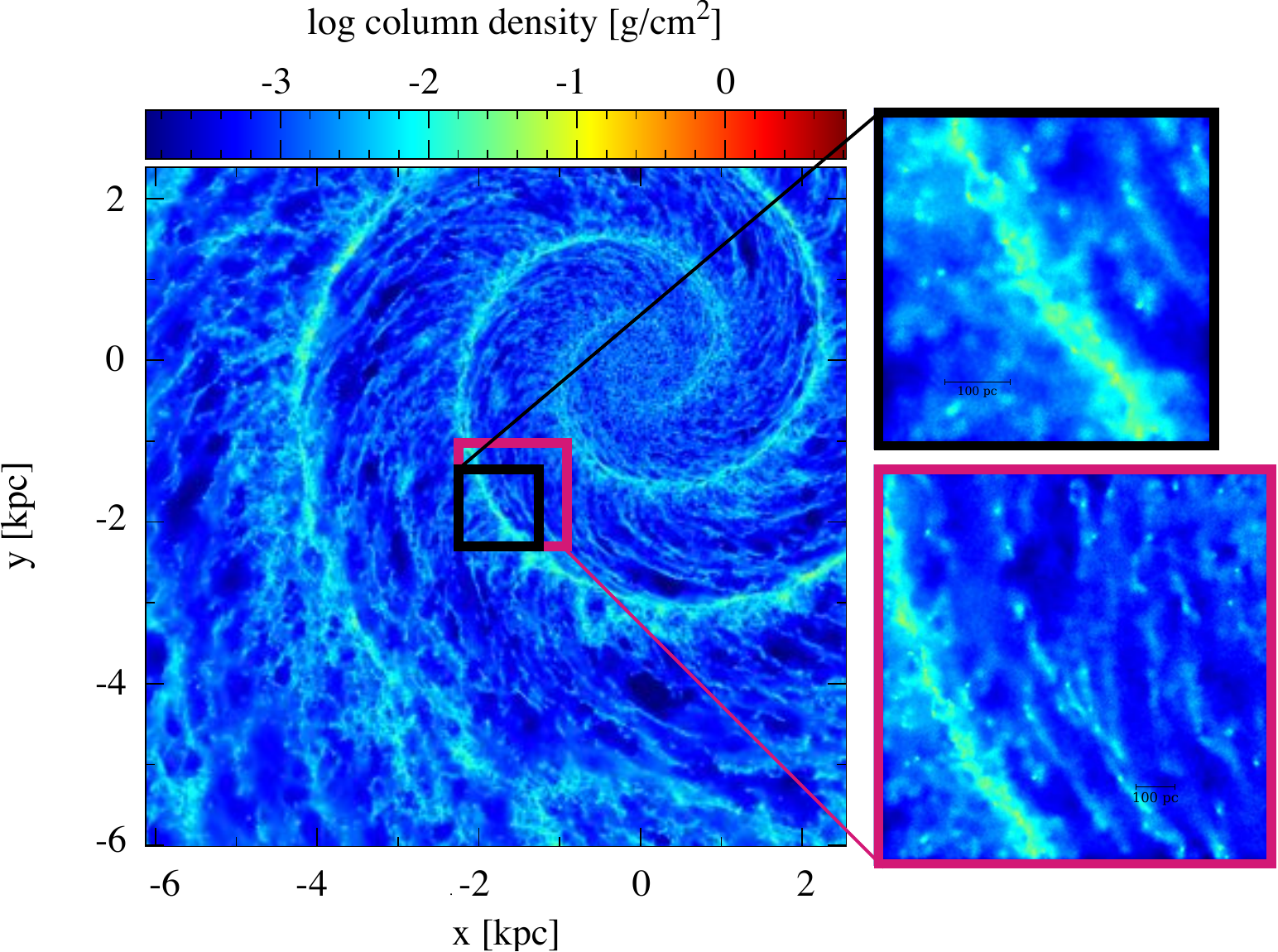}  
\end{tabular}

\caption{Spiral arm galaxy models from \citet{DobbsPringSpiral2013} in the top panel and \citet{Dobbs2017} in the bottom panel. The black box in the top panel shows the region SR1 and the purple box shows SR1e. In the bottom panel the black box shows SR2 and the purple box is SR2e. Explanations of SR1, SR1e, SR2 and SR2e are located in section \ref{section:region_extractions_and_testing} and \ref{section:model_descriptions}.}
\vspace{0.1in}
\label{fig:sr1_sr1e_sr2_sr2e} 
\end{figure}
This work uses the three dimensional smoothed particle magnetohydrodynamics code named sphNG
\citep{benz1990,benzetal1990,batebonprice1995,pricemon2007}
The code includes gravity, and ISM chemistry for heating and cooling following \citet{glover2007}. Artificial viscosity is included to capture shocks, using a switch \citep{Morris1997} such that the viscosity parameter $\alpha$ varies between 0.1 and 1. The code includes magnetic fields using a divergence cleaning method \citep{price2004a,price2004b,price2012,tricco2012,tricco2013,Dobbs2016}.

\subsection{Magnetohydrodynamics}
 The code can include non-ideal MHD physics, but for this work we assume ideal MHD. The divergence free constraint is enforced using constrained hyperbolic divergence cleaning \citep{tricco2012}, based on the method by \citet{Dedner2002}. The magnetic field is coupled to a scalar field, and the divergence error is removed from the magnetic field by propagating $\mathbf{\nabla} \cdot \mathbf{B}$ as a damped wave. We adopt the default parameters in SPMHD, and take the cleaning wave speed $c_h$ to be the fast MHD wave speed.
We find this is sufficient to keep $h \nabla . \mathbf{B}/|\mathbf{B}|$ typically $<0.01$. We also include photoionising feedback, modelled in a similar method to \cite{Dale2007b_fb} with modifications by \citet{Bending2020}. Our models contain a prescription for supernova feedback based on 
\citet{dobbs2011_sn} (see Section \ref{section:feedback_method}).

\begin{figure*} 
\begin{tabular}{cc}
\includegraphics[width=0.46\textwidth]{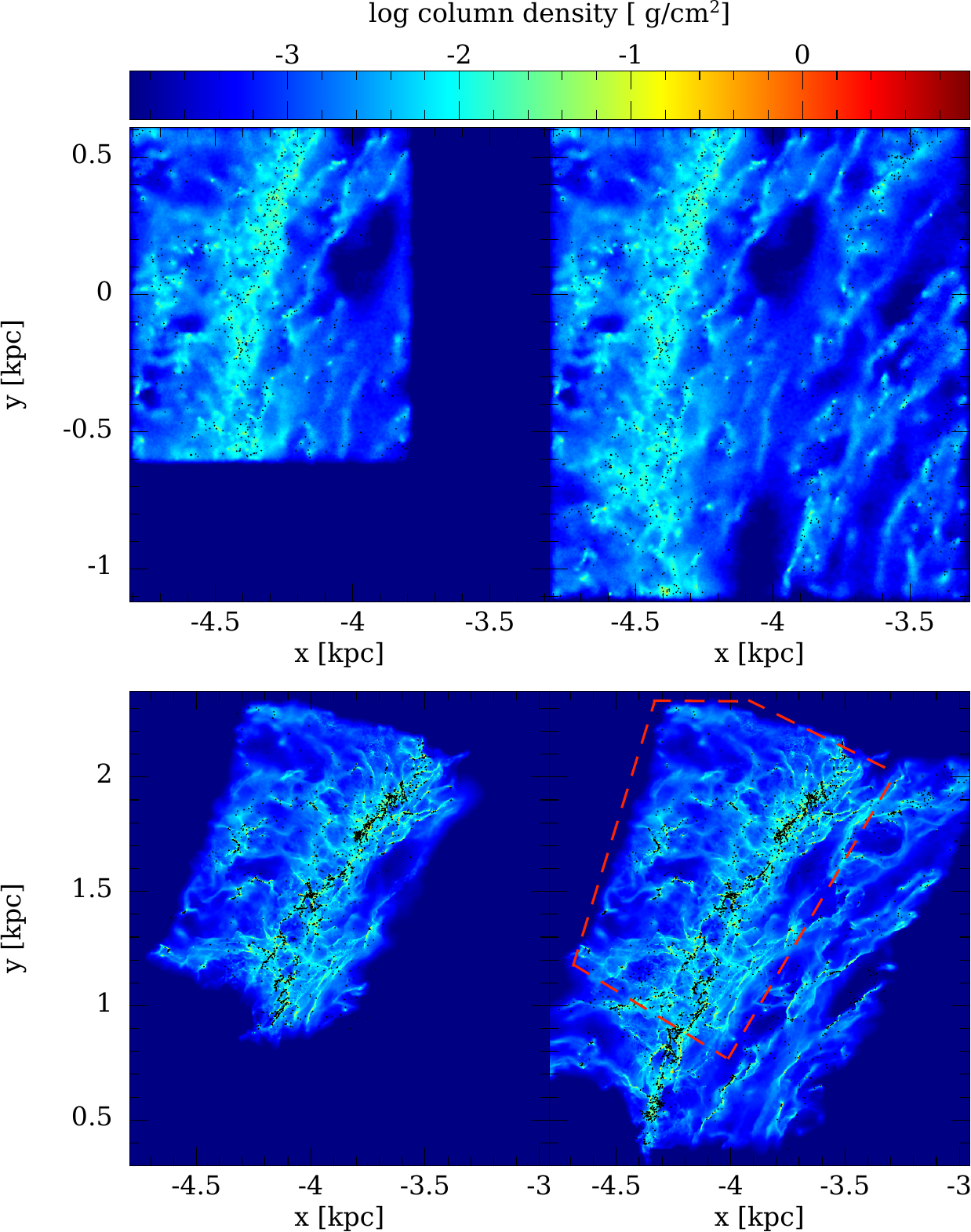}  &
\includegraphics[width=0.52\textwidth]{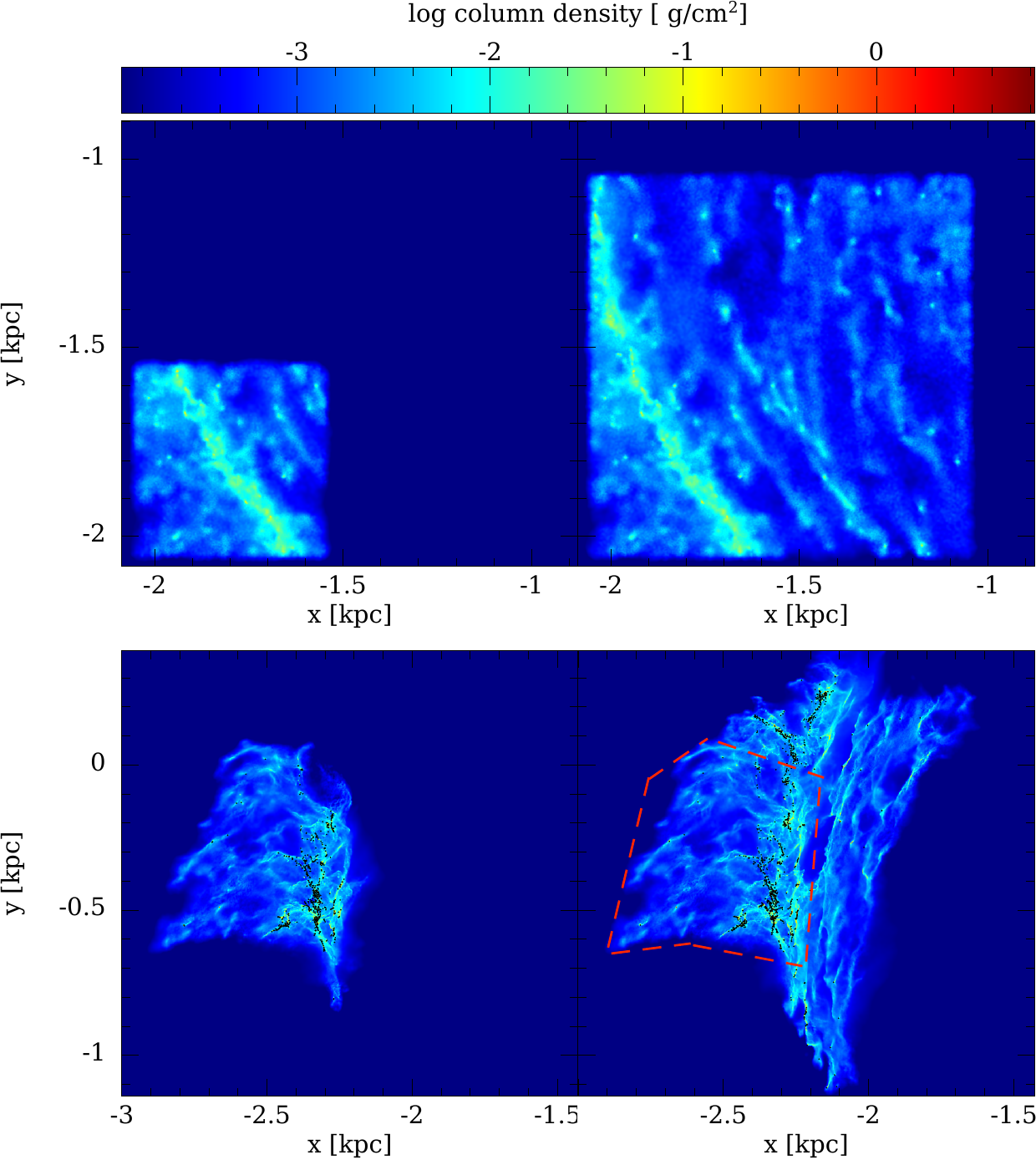}  
\end{tabular}
    \caption{Column density snapshots of models SR1 and SR1e on the left window and SR2 along with SR2e on the right window. The snapshots are taken at times t = 0 Myr and 7.7 Myrs. The equivalent sections of gas to SR1 and SR2 are shown by the red dotted lines on the bottom row of SR1e and SR2e. These regions within the red dashed lines are referred to as SR1e(SR1) and SR2e(SR2) in the main text.}
    \label{fig:region_testing}
\end{figure*}
\label{section:sink_particles}

In our simulations we do not resolve individual stars. We use cluster sink particles which represent clusters (or sub-clusters) and their contained gas. The method is described in detail in \citet{Bending2020}, here we provide a summary and highlight some changes. The cluster sink particles form according to the sink criteria detailed by \cite{Bate1995sinks} and we set the ratio of stars to gas at 0.25. In \citet{Bending2020} we used star to gas ratios of 0.5 and 1 and found that these produced high star formation rates and levels of feedback. Previous observation and theoretical work also suggests star formation efficiencies of $\sim$ 0.3 on scales below our resolution, for example jets and outflows \citep{MatznerMcKee2000,Alves2007,Federrath_etal2014,OffnerChaban2017,Pezzuto_etal2021}. 
Once a cluster sink particle is formed it can accrete gas within a given accretion radius. In this work we set this to 0.6 pc.

Stars are assigned to cluster sink particles from a list that is pre-sampled from a Kroupa IMF \citep{kroupa2001}. The sample limits are $0.01 M_\odot$ to $100 M_\odot$, however, we discard all non-massive stars ($< 8 M_\odot$) since their photoionising feedback is negligible and they do not undergo SNe. In \citet{Bending2020} and \citet{bending2022} the limit was $18 M_\odot$, however, stars that undergo SNe on timescales longer than 10 Myr now become relevant with the inclusion of an initial population (Section \ref{section:initial_population_of_stars}). We continue to only include the photoionising flux from stars with mass greater than $18 M_\odot$. 

Massive stars are added to the simulation each time the total stellar mass in cluster sink particles increases by $\Delta M_{i}$ (the massive star injection interval). $\Delta M_{i}$ is calculated from the pre-sampled list as the total mass of the sample over the number of massive stars in the sample; in this work this gives 305 M$_{\odot}$.
The properties of each massive star (mass, Lyman flux, lifetime) are given from the pre-sampled list.
Each massive star is allocated to the sink with the largest non-stellar mass. This non-stellar mass is calculated as $M_{\textrm{sink}}\times 0.25-N_{\textrm{massive}}\times \Delta M_{i}$ where $M_{\textrm{sink}}$ is the mass of the sink, 0.25 is the fraction of mass assumed to be stars (see above) and $N_{\textrm{massive}}$ is the number of massive stars in the sink (i.e. this represents the total sink mass minus the mass which has already been accounted for as contributing massive stars).
This is a slight modification from \citet{Bending2020} in which the sum of the mass of massive stars in each sink was used in the calculation rather than $N_{\textrm{massive}}\times\Delta M_{i}$, this meant larger sinks contained disproportionately large numbers of ionising stars. 

\begin{table*}
\caption{Descriptions of the initial conditions of all the models in this study. Each model has a mass resolution of 3.68 \Ms per particle.} 
\begin{tabular}{m{7em} m{7em} m{7em} m{6em} m{5em} m{8em} m{8em}}

\hline
 & Name & Initial & Feedback & MHD & Size of   & Number \\
 & & Population & & & region (kpc$^{2}$) & of Particles\\
 \hline
   \multicolumn{7}{l}{Region extraction testing}\\
   \hline
 &SR1 & yes & none & none & 1.20 & $4.3 \times 10^6$\\ [2ex]
 &SR1e & yes & none & none & 2.55 & $7.2 \times 10^6$\\ [2ex]
 &SR2 & no & none & none & 0.50 & $1.1 \times 10^6$\\ [2ex]
 &SR2e & no & none & none & 1.00 & $2.4 \times 10^6$\\
 \hline
   \multicolumn{7}{l}{Feedback and MHD testing (using SR1)} \\
   \hline
 &NOFB & yes & none & none & 1.20 & $4.3 \times 10^6$\\ [1ex]
 &IOFB & yes & Photoionising & none & 1.20 & $4.3 \times 10^6$\\ [2ex]
 &IOSFB & yes & photoionising \& supernovae   & none & 1.20 & $4.3 \times 10^6$\\ [2ex]
 &IOSFBc & no & photoionising \& supernovae   & none & 1.20 & $4.3 \times 10^6$\\ [2ex]
 &MHD5 & yes & none & B$_{\textrm{ini}}=5$ µG & 1.20 & $4.3 \times 10^6$\\ [2ex]
 
 \hline   
\end{tabular}
\label{tab:models}
\end{table*}

\begin{figure}
    \centering
    \includegraphics[width=83mm]{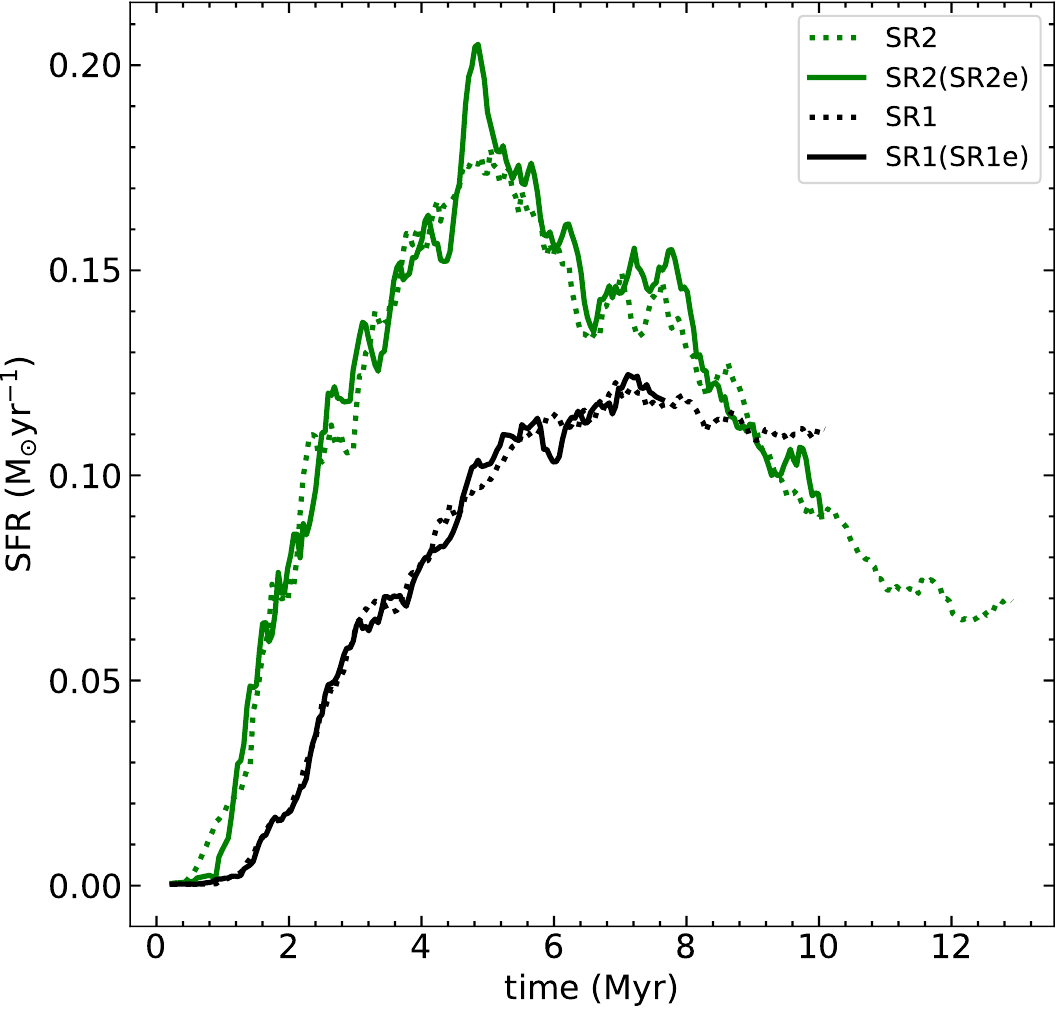}
    \caption{
    The star formation rates of models SR1 and SR2 (solid lines) are compared to models SR1e and SR2e. The dotted lines, referring to SR1e(SR1) and SR2e(SR2), show the star formation rate taken from an equivalent section of gas to SR1 and SR2 within the extended models (see the red dotted regions in figure 2). The star formation rate in the equivalent regions are very similar, whether the extended surroundings are included or not.}
    \label{fig:sfr_region_testing}
\end{figure}

\subsection{Photoionising and supernova feedback}
\label{section:feedback_method}

This work includes two forms of stellar feedback, photoionisation and supernovae. The photoionisation method follows a line of sight (LOS) approach described in \citet{Bending2020} and the supernovae are modelled as described in \citet{dobbs2011_sn} and \citet{bending2022}. Here we will briefly describe the two methods and some improvements.

For the photoionisation a LOS is drawn between every particle-sink pair. The gas particles for which the LOS intersects their volume of compact support are identified. The column density contributions are calculated by a line integral through the smoothing kernel of each of these particles, which is integrated to get a contribution to the column density. These contributions are summed to get the total column density for each line of sight. Once we have computed the column densities we use the on-the-spot approximation to treat the direct to ground state recombinations. 

We have improved the way in which overlapping HII regions are dealt with since \citet{Bending2020}. The issue is that gas particles can receive ionising flux from more than one cluster sink particle leading to ionising flux from a smaller or more distant source being discarded. We have addressed this using a similar approach to \citet{dale2011}. They track the number of sources contributing to the ionisation of each gas particle and then iterate reducing the column density of those particles by a factor of the number of sources until they reach some tolerance. This iterative approach is highly computationally expensive so we instead use the number of sources calculated in the previous timestep to reduce the column densities. \citet{dale2012} improved the method to consider the fractional contributions of each source to the ionisation of any given gas particle, however, they found this only had a very small effect. We have not implemented this since it has a high memory requirement.

\begin{figure*}
    \centering
    \includegraphics[width=175mm]{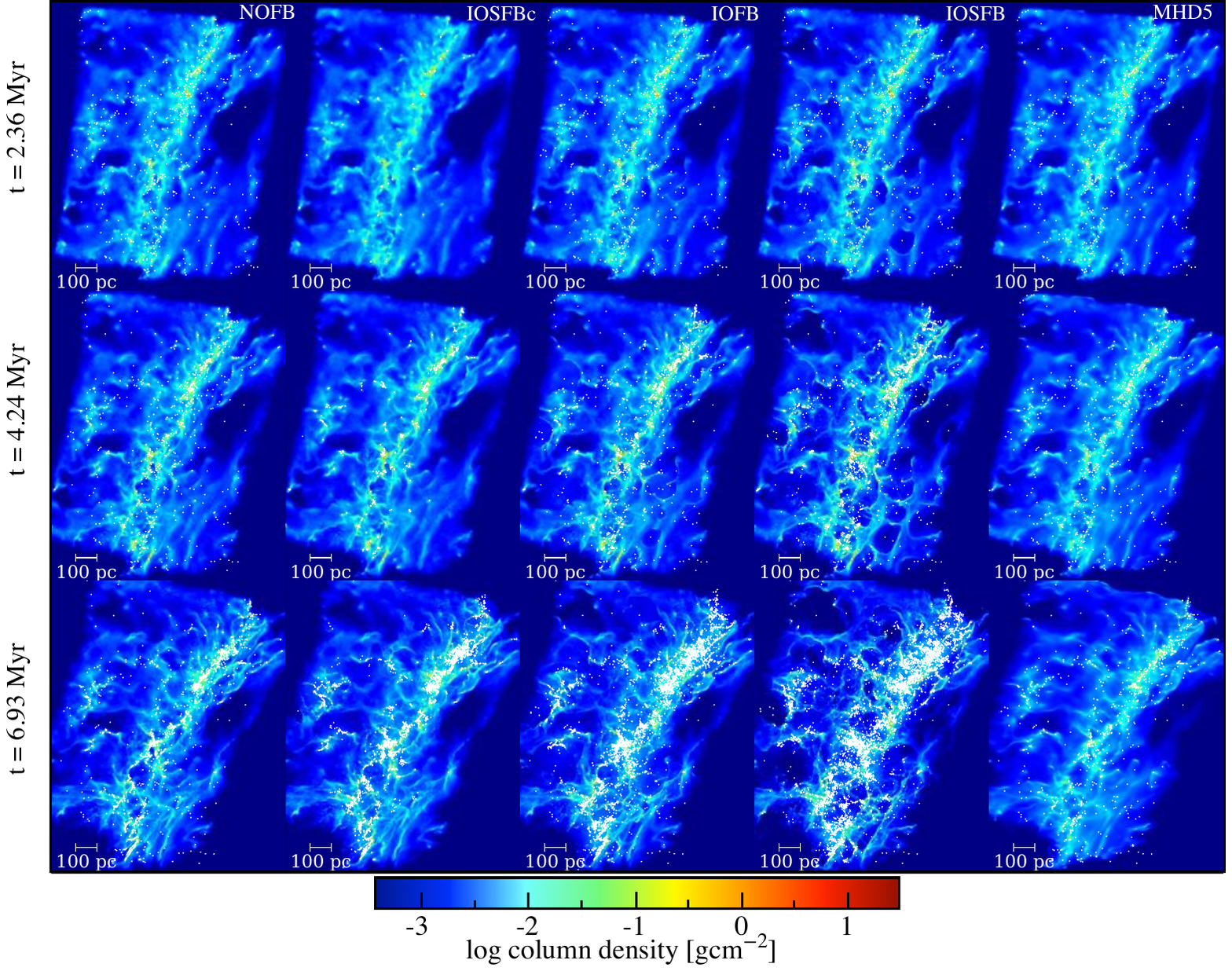}
    \caption{Column density snapshots of all models going left to right NOFB, IOSFBc, IOFB, IOSFB and MHD5 at 2.36, 4.24 and 6.93 Myr. A single column represents a model with each row displaying a snapshot in time. Cluster sink particles are displayed in each snapshot as white scatter points.}
    \label{fig:allmodels}
\end{figure*}

Supernovae are modelled using the same method as described in \cite{dobbs2011_sn} and \cite{Bending2020}, by inserting kinetic and thermal energy following the snowplough solution. Stellar lifetimes are computed using the Seba code \citep{Portegies2012} and once a massive star is inserted into a cluster sink, the sink's current age is checked against the lifetime of the massive star. When the age of the sink exceeds the massive star lifetime, a supernova is initiated and photoionization ceases.

\subsection{Initial Conditions}
In this paper we perform a study of how sensitive our simulations are to the initial conditions. We consider two aspects of the wider galaxy scale physics, to probe their effects on small scale star formation. First we test whether extracting a small region from a large scale simulation is reliable, i.e. do we see the same results in the small region compared with what would be found in a large scale simulation. Next we look at whether prior star formation acting on galaxy scales affects smaller scales by including an initial population of stars. So effectively in the former we are testing the spatial dependence of the initial conditions, whilst in the latter we test the temporal dependence of the initial conditions. These initial condition tests are described in sections \ref{section:region_extractions_and_testing} and \ref{section:initial_population_of_stars} respectively. 
\subsubsection{Region extraction}
\label{section:region_extractions_and_testing}
The initial conditions for our models are taken from two spiral galaxy simulations from \cite{DobbsPringSpiral2013} (GM) and \cite{Dobbs2017} (GMs). These two galaxy models start from the same initial conditions and include self-gravity, 
a galactic potential, ISM chemistry for heating and cooling and supernovae feedback. The only difference is that GMs also includes star particles (see section \ref{section:initial_population_of_stars} for more details on star particles). 
Once our region is chosen we use a resolution increase method detailed by \cite{Bending2020}, in which particles are arranged around a spherical grid centred on the target particle. The radius of the spherical grid is $2h$, where $h$ is the smoothing length. The particles are distributed over a spherical grid according to the SPH smoothing kernel. In the parent galactic models the mass per particle is 312.5 \Ms, after splitting the particles using the resolution increase method we achieve 3.68 \Ms per particle. This provides us with enough resolution to model ionising feedback effectively, without increasing the particle numbers such that the models evolve on unreasonable time scales.

We simulate two different regions, which we call SR1 and SR2, both of which are box-shaped. To test the region extraction, we performed a number of tests, where we extended the SR1 and SR2 in different ways, but focus on two models, SR1e and SR2e here.
The region extraction models are run without feedback, since these are faster, and we are simply testing whether the size of the region makes a difference, which does not depend on feedback. 
The first selected region and its extended version will be referred to as SR1 and SR1e throughout. This region is taken from the GMs model at a radius of 4 kpc from the galactic centre, visible in the top panel of figure \ref{fig:sr1_sr1e_sr2_sr2e}.
 SR2 is the same region used in the \cite{Bending2020} models taken from GM (the same galactic model used in their work), residing roughly 2.8 kpc away from the galactic centre, shown in the bottom panel of figure \ref{fig:sr1_sr1e_sr2_sr2e} along with the extended model SR2e. We extend both SR1 and SR2 by 0.5 kpc in the (x, -y) and (x, y) plane respectively. 
\subsubsection{The initial population of stars}
\label{section:initial_population_of_stars}
Our models containing an initial population of stars are extracted from the galaxy model GMs (i.e. SR1 described in the previous section). This model contained star particles which represent sites of star formation. These particles are fixed mass and non accreting with ages between 0 and 40 Myr. The age and mass of each star particle within the region of choice are carried over to the models presented here.
To 
model
feedback we convert the star particles from GMs into cluster sink particles as described in Section \ref{section:sink_particles}. This means that the star particles are assigned a stellar population when creating the initial conditions, and some will contain massive stars which will immediately undergo ionisation. Some will also imminently undergo supernovae. The sink particles can also undergo accretion which was not the case for the star particles in the GMs simulation.

\begin{figure}
    \centering
    \includegraphics[width=84.5mm]{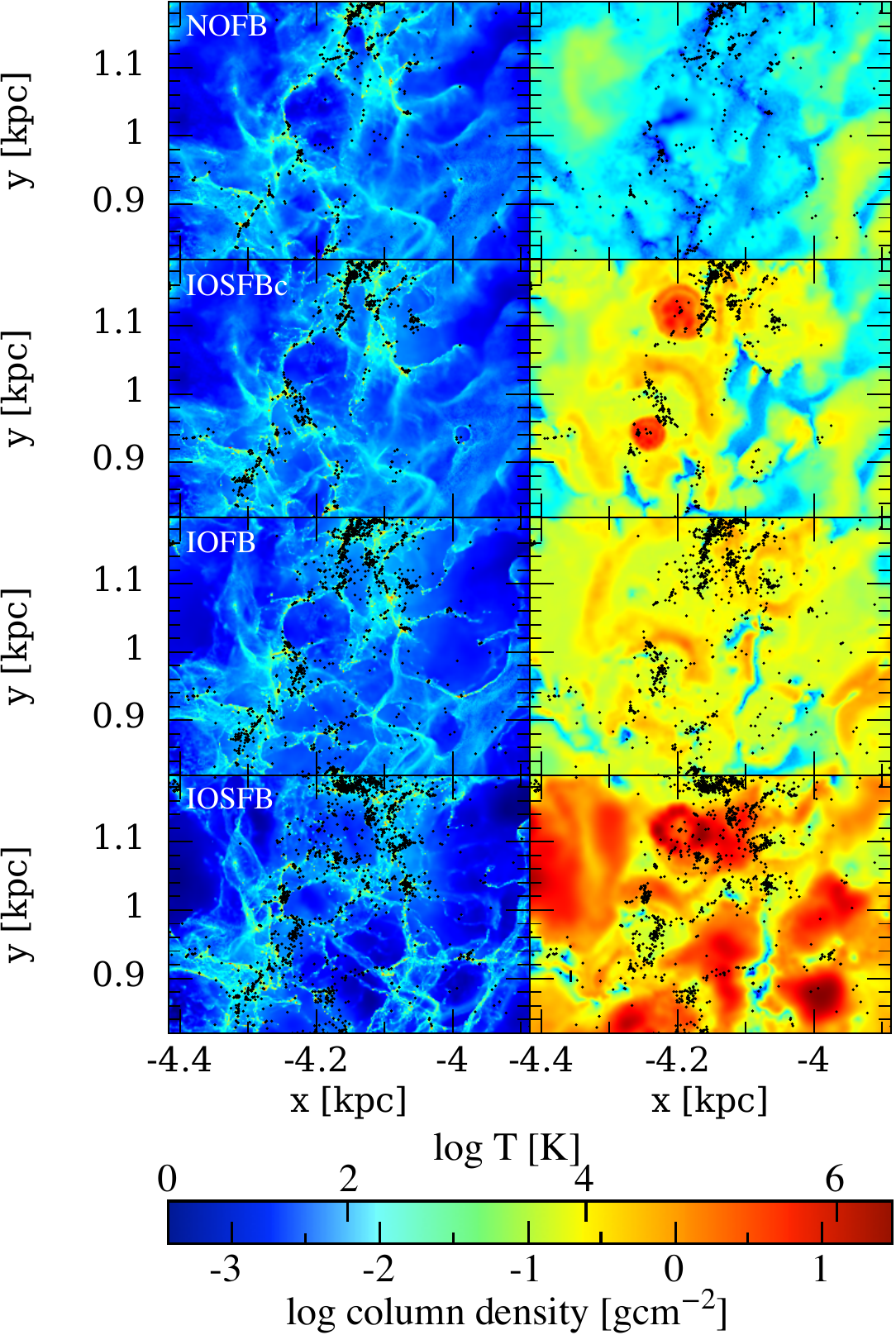}
    \caption{Feedback models IOSFB and IOFB are shown in comparison with our two control runs NOFB and IOSFBc. These snapshots which feature a zoomed in region in the lower half of the central spiral arm, show column density in the first column and temperature in the second column, at time 6.93 Myr. Cluster sink particles are shown as black points.}
    \label{fig:CDandTzoom}
\end{figure}

\subsubsection{Magnetic field}
For one of our models, we set up a galactic toroidal magnetic field, with an initial magnetic field strength of 5 $\mu$G. This field strength is consistent with observations of magnetic field strengths in spiral galaxies and the solar neighbourhood \citep{Sofue86,Rand89,Ferriere01,Beck15,Han17,Patel_review_2022}. We did not include feedback in this model, rather we use this model to see how much magnetic fields affect star formation compared to the no feedback model, and how the addition of magnetic fields compares to sensitivity to initial conditions. 

\subsection{Model descriptions}
\label{section:model_descriptions}
We list the models in Table \ref{tab:models}. The first set of simulations are the region extraction tests, as described in Section \ref{section:region_extractions_and_testing}. These all have the same resolution, but cover larger and smaller regions of the galaxy.
The next set of models test feedback from an initial population of stars. These models use the same region as SR1, from the galaxy model with star particles, and
vary by the feedback mechanisms included. These are split into 4 models containing two control runs. Control run one has an initial population of stars and no feedback to compare to our feedback runs. This will be referred to as NOFB (no feedback). Our second control run has no initial population of stars but includes all forms of feedback to assess the impact of the initial population of stars, which will be named IOSFBc. The remaining two runs have an initial population of stars, comparing ionising feedback only (IOFB) and ionising feedback with supernovae (IOSFB). Our final model (MHD5) is performed without feedback but setup with a galactic toroidal magnetic field, using the same region as SR1. 


\section{Results}
\subsection{Region extraction tests}

To test the region extraction method, we compare SR1 with SR1e and SR2 with SR2e. We show SR1, SR2, SR1e and SR2e, all without feedback at a time of 7.7 Myr in Figure~\ref{fig:region_testing} (bottom panel). To aid the comparison we have drawn the equivalent region of gas for SR1 and SR2 within the extended models, shown by the dotted red line in figure \ref{fig:region_testing}. For SR1 (left panels), we see minimal difference between the isolated region SR1 versus the equivalent part of the extended model SR1e. This implies that taking SR1 only is a reasonable representation of how the region would evolve within the host galaxy throughout our simulation time frame.

For SR2 (right hand panels) we see that there are significant differences between the SR2 region and the equivalent area of gas in the extended SR2e model. These are apparent at for example the top of the SR2 region where there is a large(r) cavity or hole and fewer sinks. There is also more gas apparent in the spiral arm, the features are slightly sharper in SR2, and the distribution of sinks is slightly different. 


The difference between the two cases arises from how well the gas flow into the region is captured. For both regions, gas is flowing from right to left, entering roughly perpendicular to the spiral arm. For SR1, the arm is situated to the left, there are several 100's parsecs of gas to flow into the arm along all its length, and even at the later time, there is still some gas which has yet to enter the spiral arm. For SR2, the geometry of the region is different, such that at the top of the spiral arm for this region, there is no gas present (i.e. perpendicular to the arm) to flow into the arm. Furthermore at the later time, there is no or little gas still to enter the spiral arm. The only place where there is much gas which can flow into the arm is in the middle, i.e. where gas flowing in extends up to the top right corner. These difficulties are exacerbated at smaller galactic radii, since the difference between the pattern speed of the spiral and the rotation of the galaxy is higher. Taking the velocity of the gas relative to the spiral arm, we estimate that the timescale for gas to travel into the spiral arm in SR1 is similar to the length of the simulation ($\sim$ 10 Myr), whereas in SR2 the timescale is shorter. Thus the simulation time, coupled with the velocity of the gas relative to the spiral arms, give an indication of the required size of the region which can reasonably be simulated.
Another difference, as well as gas replenishment, is that gas at the boundaries will also add additional pressure terms at the edge of the region.

We checked the star formation rates for the models, where we would expect that lack of gas replenishment would mean lower star formation rates. We calculate the star formation rates for each simulation, selecting only those particles in SR1e and SR2e which are also in SR1 and SR2, shown in figure \ref{fig:sfr_region_testing}. We found that the difference in star formation rates between the extended models versus SR1 and SR2 was surprisingly small. We find the star formation rates over the same particles in SR1e and SR1 are near identical. For the second region, we see a slightly higher star formation rate in SR2e compared to SR2, but overall quite similar. 

We tested extending SR1 and SR2 in directions both above and below the original regions by 0.5 kpc and found that further extension in the $y$ direction (i.e. along the spiral arm) made little difference to the results obtained with SR1e and SR2e (i.e minimal difference in star formation, similar morphologies of gas in the red outlined region of interest in Figure~\ref{fig:sfr_region_testing}) suggesting that the extension in these directions is less significant than perpendicular to the arm.




\subsection{Region evolution with initial population}

In the previous section we looked at the spatial extent of the initial  conditions, here we consider the effect of star formation history by comparing with and without a prior population of stars. In the absence of feedback, an initial population of stars will make minimal difference, so here we are investigating feedback from previously formed stars. The following simulations are performed with the initial conditions for the SR1 model from our previous tests.

Figure~ \ref{fig:allmodels} shows column densities for the models where we test the impact of an initial population, at 3 different times. At the earlier time of 2.36 Myr, there are minimal differences between the models. At 4.24 Myr we start to see more structure in the IOSFB model, and by 6.93 Myr there are clear if not large differences between the simulations. The largest difference by eye in terms of structure for the non-MHD models is between the NOFB and IOSFB cases, where we see clearer structures, and there more sink particles in model IOSFB. At 6.93 Myr NOFB has formed 2789 sinks where as in IOSFB there are 8103 sinks. In all our models, as our regions evolve, the molecular clouds along the spiral arm and inter-arm regions begin to collapse and star formation initiates within the first million years. With feedback, but no initial population (IOSFBc), photoionisation and supernovae will start to shape the gas from the newly formed sinks, but even by 6.93 Myr this has not had a large impact on the gas structure compared to the no feedback model (NOFB). However for the IOFB and IOSFB models, there is both additional feedback from the initial population, and this feedback occurs at an earlier time. We see that IOSFB has the most complex structure, with HII regions, and sharp filaments, then IOFB has less followed by IOSFBc and NOFB. There  are also 
correspondingly fewer sink particles in the models with no initial stellar population and fewer feedback processes.    

Thus it appears in our models that increasing the feedback increases the structure in the gas, and at least from the number and distribution of sink particles, also appears to increase star formation. This is most evident in the simulation with an initial population of stars which undergo both ionisation and supernovae, which thus contains both feedback processes operating from the outset of the simulation. From the outset, these are shaping the gas. Dense filaments along the spiral arm are compressed by the expansion of HII regions, whilst supernovae sweep gas through low density regions impacting dense material further aiding collapse. 
 
For the MHD model (MHD5), the effect is the opposite (to increasing feedback), the magnetic fields appear to smooth out the structure compared to the other models including NOFB. This is due to magnetic pressure preventing the formation of such dense structures.

Figure \ref{fig:CDandTzoom} shows a zoom in of models NOFB, IOSFBc, IOFB and IOSFB, with column density and temperature along columns 1 and 2 respectively. With no feedback the region is cold reflecting the temperature of the cold neutral medium and molecular cloud phases of the ISM. Including ionising feedback results in bubbles around sinks that are heated to around $10^4$ K leading to more warm ionised medium in the ISM. Adding supernova feedback to this means that HII regions are further heated up to roughly $10^6$ K. 
Looking at the temperature in the second column of figure \ref{fig:CDandTzoom} it is clear that newly forming sinks are distributed around the edges of feedback bubbles. 
Comparing our control run IOSFBc with IOFB we see that despite IOSFBc having both supernova and ionising feedback the column densities of these two models are relatively similar, at least IOFB is more comparable to IOSFBc compared to IOSFB. We see some difference in the temperature panels, with hot gas due to two recent supernovae evident. Again if we compare the lower panel, with IOSFB, we see many more supernovae have occurred which is having a greater impact on the density.
\subsection{Star formation}
\label{section:sfr_sfe}
\begin{figure}
    \centering
    \includegraphics[width=82mm]{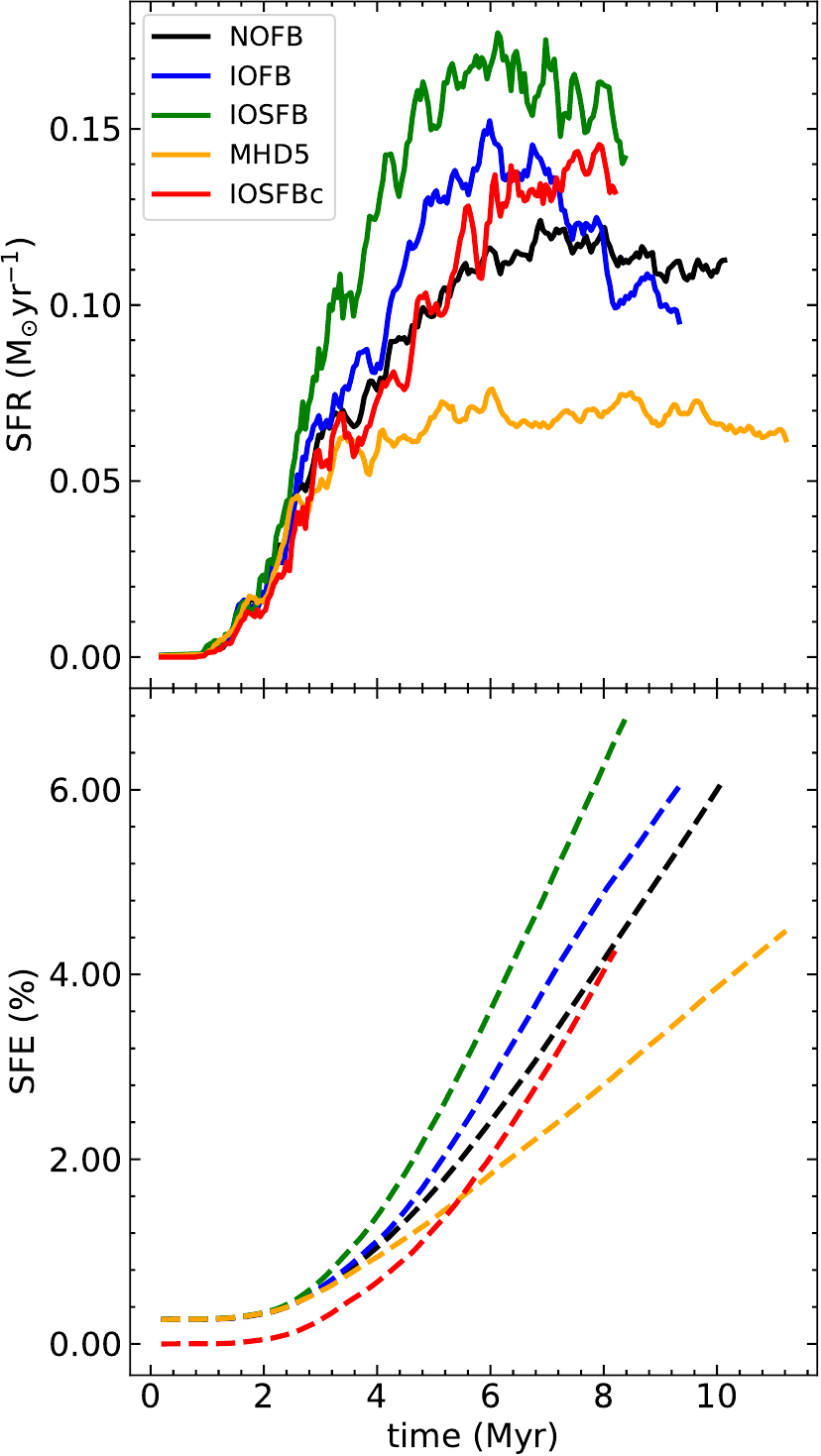}
    \caption{Star formation rate is shown in the top panel and star formation efficiency in the bottom panel for models NOFB, IOSFBc, IOFB, IOSFB and MHD5. The details of these calculations are explained in section \ref{section:sfr_sfe}.}
    \label{fig:sfr_sfe}
\end{figure}
We present the star formation efficiency (SFE) and star formation rate (SFR) in figure \ref{fig:sfr_sfe}. They are calculated by the relations
\begin{equation}
    \textrm{SFE} = \frac{0.25\times M_*}{0.25\times M_* + M_{\textrm{gas}}}
\end{equation} 
and
\begin{equation}
    \textrm{SFR}(t_i) = 0.25 \times \frac{M_{*}(t_i) - M_{*}(t_{i-1})}{t_{i} - t_{i-1}}
\end{equation}

\begin{figure*}
    \centering
    \includegraphics[width=183mm]{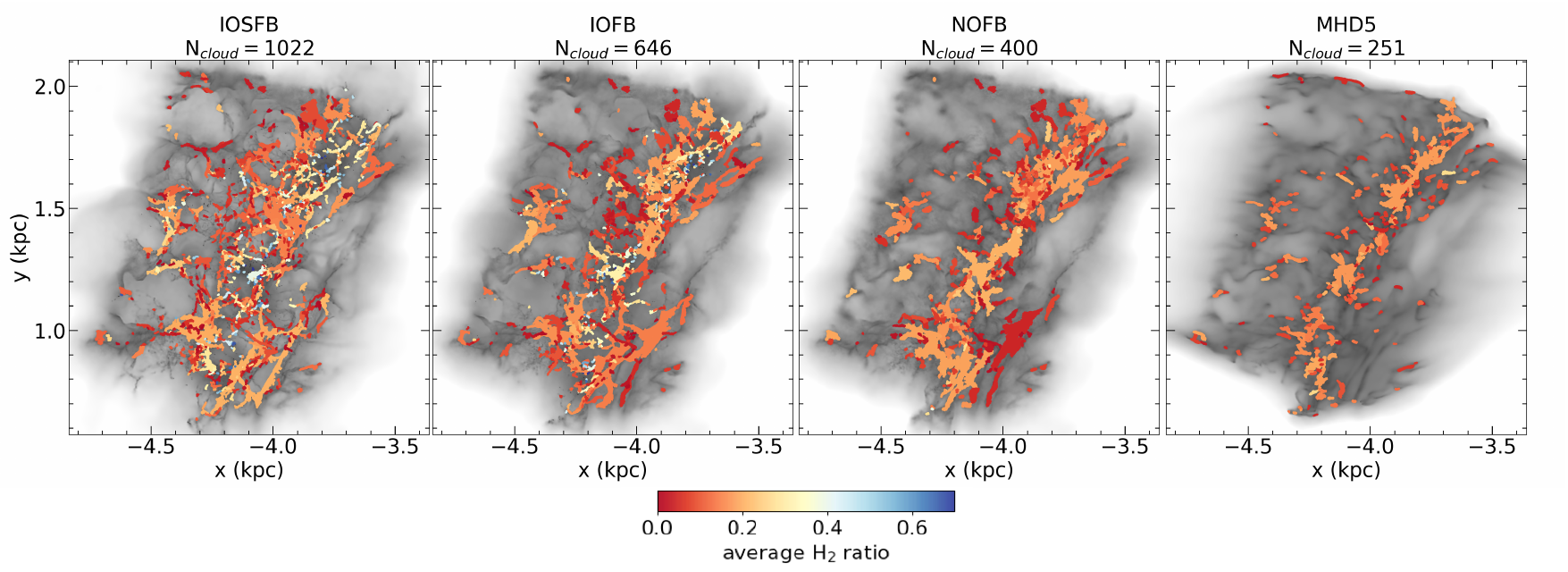}
    \caption{Snapshot of models NOFB, MHD5, IOFB and IOSFB are shown at 6.6 Myr. The column density is rendered in grey with the clouds plotted on top coloured by their average ratio of H$_2$.}
    \label{fig:cloud_h2}
\end{figure*}

where $M_* = \sum_{n=1}^{N_{\textrm{sink}}} m_{\textrm{sink},n}$, $M_{\textrm{gas}} = \sum_{n=1}^{N_{\textrm{gas}}} m_{\textrm{gas},n}$, 0.25 is the fraction of mass in sinks which is assumed to be converted into stars and $t_{i} - t_{i-1}$ refers to the difference in simulation time between two neighbouring data dumps. Our models with feedback and an initial population of stars (IOSFB and IOFB) begin to diverge in SFE (lower panel of figure \ref{fig:sfr_sfe}) from the no feedback case (NOFB) after 3 Myr. At 7.00 Myr our feedback models, IOSFB and IOFB reached a SFE of 4.99 \% and 3.90 \% respectively, whereas our NOFB model reached an SFE of 3.20 \%. The control model IOSFBc starts with a SFE of 0 since there is no initial population. This continues for the first million years until star formation occurs. At this point the SFE begins to rise at a rate greater than our NOFB model, reaching 2.91 \% by 7.00 Myr. MHD5 shows a flatter slope compared to the other models, reaching a SFE of 2.27 \% by 7.00 Myr. This further shows that including both forms of feedback boosts star formation, whilst MHD suppresses star formation.

Moving to the top panel of figure \ref{fig:sfr_sfe}, we see that across all models the rate of star formation remains roughly constant for the first million years, despite the presence of an active population of stars. Before the first stars form, the star formation rate is controlled solely by the accretion of the initial population which as we show later in figure \ref{fig:snkm} is a small contribution. After 1 Myr star formation begins to increase in all cases. This is where the models diverge in SFR, as feedback alters the environment of star formation sites across the spiral arm. With all forms of feedback and an initial population of stars (IOSFB) the star formation rate increases rapidly, peaking at just below 0.2 M$_{\odot}$yr$^{-1}$ by 7.00 Myr. Our model with ionising feedback only, IOFB, (and an initial population of stars) follows the same trend but does not increase as steeply reaching a peak of around 0.15 M$_{\odot}$yr$^{-1}$ by the same time. This burst of star formation with ionising feedback is similar to models by \citet{Bending2020}. When we remove the initial population but include all forms of feedback (IOSFBc) we find a similar shape to the star formation rate but with a stunted peak. This shows the importance of the early supernovae activity in boosting the star formation rate. 

Finally without any feedback and with an initial population of stars (NOFB) we see the star formation rate increase as gas collapses into stars, and flatten out over time, in the same way as seen by \citet{Bending2020} and \citet{Ali2022}. However including magnetic fields suppresses the peak of the star formation rate. Following the MHD5 model the SFR increases in the same way as our other models but plateaus between 3 and 4 million years at SFR values of 0.05 - 0.053 M$_{\odot}$yr$^{-1}$, which means star formation suppression by magnetic fields is more influential on the star formation rate than the inclusion of an initial population of stars.

We see that in all cases the star formation rates are higher with feedback, more with an initial population of stars, but still IOSFBc is slightly higher compared to the no feedback model. The increase reflects triggered star formation. Whilst feedback can also inhibit star formation on small scales, this depends somewhat on the sink radius. If the sink radius is smaller, accretion on to the sink is reduced, and for those sinks with feedback, feedback instead heats the surrounding gas so that it doesn’t participate in star formation. So for example in \citet{Bending2020}, the sink radii were larger and the increase in star formation greater compared to this paper, whilst with sufficiently small sink radii the feedback produces a net decrease in star formation \citep{dobbs2022}. The impact of feedback from the initial population however will be predominantly to trigger star formation since accretion onto these sinks is minimal (see Section \ref{section:impact_of_ini_pop}).

\subsection{Clouds}
\begin{figure*}
    \centering
    \includegraphics[width=163mm]{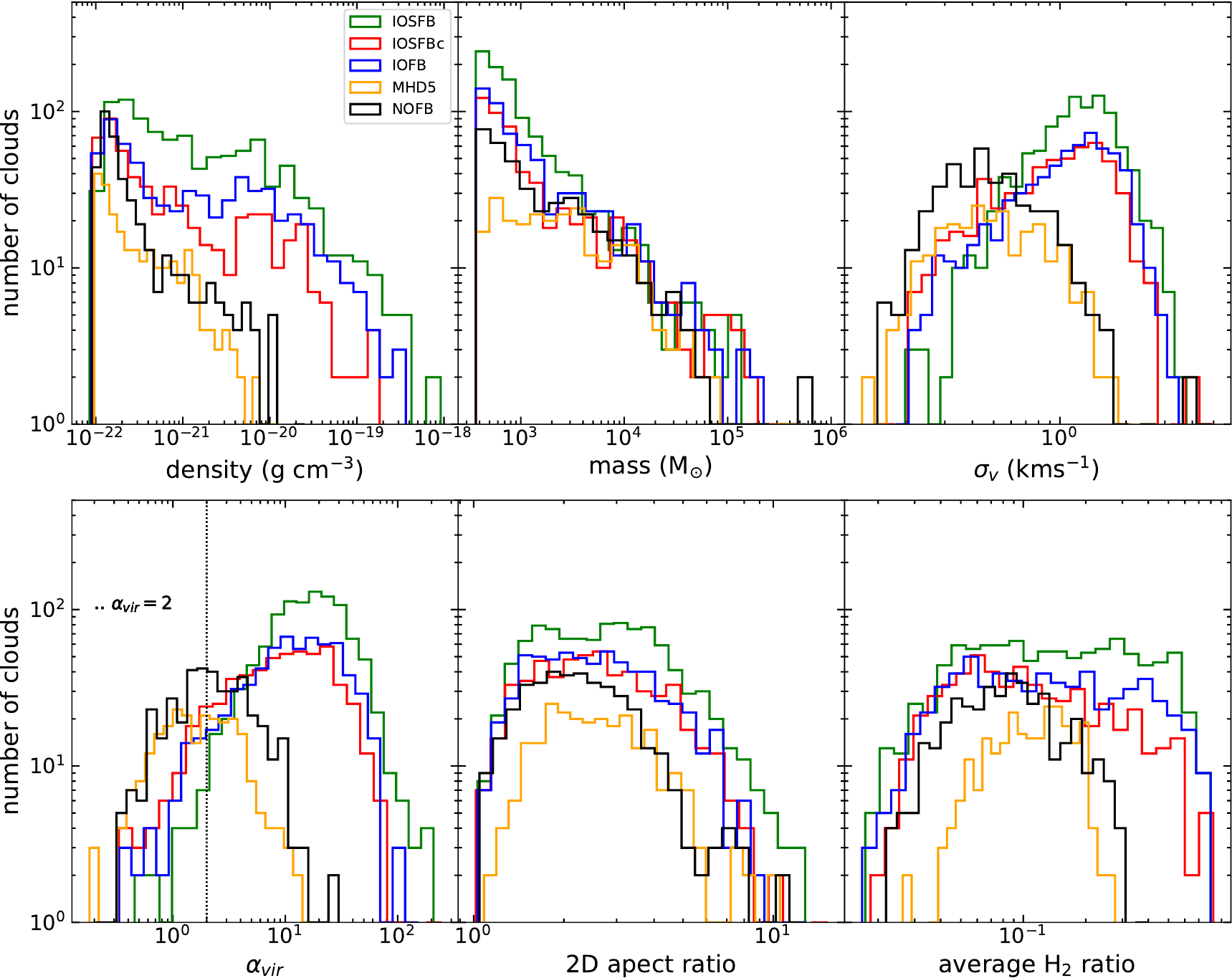}
    \caption{Properties of clouds are shown across all no feedback and feedback models. The first row shows cloud density, masses and dispersion velocities, spanning left to right. Row 2 shows the virial parameter, 2D aspect ratio and average H$_2$ ratio for all clouds (again spanning left to right).}
    \label{fig:cloudprop}
\end{figure*}
Using a friends of friends algorithm we identify cloud structures within our models, similar to \citet{Bending2020}. The algorithm groups particles together based on a separation of 1.26 pc to identify members of a cloud. A particle is chosen and grouped with all other particles which lie within 1.26 pc. The particles added to the group are checked in the same way until no more ungrouped particles are within 1.26 pc of any in the group. The process is repeated for ungrouped particles until all particles have been assigned to a group. We do not consider groups of less than 100 particles to be clouds.

We compare the molecular hydrogen content of the clouds formed in four of the models (IOSFB, IOFB, NOFB and MHD5) in figure \ref{fig:cloud_h2}. This figure shows the column density of each
region
with the clouds overlaid on top. The colour scale of the clouds shows the average H$_2$ ratio. The H$_2$ ratios will be a lower estimate due to limitations of resolution \citep{Duarte-Cabral2015,Joshi2019}.
The number of clouds found follows the same trend as \cite{Bending2020} and \cite{Ali2022} in which adding feedback leads to cloud fragmentation and so increases the number of cloud structures across the region. We also find that adding MHD to our models reduces the number of clouds significantly. At 6.6 Myr, with MHD we find 250 clouds (MHD5), NOFB 399 clouds, 
IOFB 645 clouds and IOSFB 1021 clouds. Feedback is leading to the fragmentation of clouds, resulting in higher numbers of lower mass clouds across the region. 
Feedback appears to break up the clouds presumably by increasing density variations, whereas magnetic fields essentially smooth the gas and reduce fragmentation. 
We also see that in the models where feedback is having more of an effect, i.e. IOSFB and IOFB with the initial population, the molecular gas ratios are higher. As we see in the next paragraph feedback is acting to compress gas to higher densities than the gas would be otherwise, and consequently higher H$_2$ ratios.

In figure \ref{fig:cloudprop} we compute the properties of the clouds in the different simulations. Starting with the density, our feedback models IOSFB, IOSFBc and IOFB show a larger number of clouds with higher densities, indicating that 
clouds in these models are compressed by the action of feedback.
The mass distribution of the clouds is similar for each model, though the number of low mass clouds is suppressed with magnetic fields, perhaps because some low mass structures are smoothed out and no longer meet the criteria for a cloud.

We see that the majority of clouds have larger  velocity dispersions in models IOSFB, IOSFBc and IOFB, typically around 1 - 2 kms$^{-1}$ with a small number of clouds containing velocities of 4 kms$^{-1}$. 
This is due to momentum from feedback driving the gas motions.
Without feedback (NOFB and MHD5), velocity dispersions 
only reach 2 kms$^{-1}$ with most clouds containing velocities of 0.2-0.4 kms$^{-1}$. 

In the bottom left panel of figure \ref{fig:cloudprop} we calculate the virial parameters for the clouds. Most of the clouds within the feedback models are unbound, which is expected as they have higher velocity dispersions from feedback driven motions. NOFB and MHD5 again show similar properties with roughly even numbers of clouds that are bound and unbound. We do see that there is a dip in the number
of unbound clouds in the MHD5 model compared to NOFB. This may again be due to magnetic pressure smoothing out low density features.

Moving to the middle panel of the bottom row of figure \ref{fig:cloudprop}, we present the 2D aspect ratio of the clouds. Using this in conjunction with figure \ref{fig:cloud_h2}, the morphology of the clouds can be considered. Models IOSFB and IOFB show clouds with more elongated and filamentary shapes, located at the extent of feedback bubbles. The 2D aspect ratio confirms there are more clouds in the model with the most feedback occurring (IOSFB) with larger aspect ratios. Feedback appears to fragment and shape clouds across the region. The models MHD5 and NOFB in figure \ref{fig:cloud_h2} show fewer elongated structures, and contain no very high aspect ratio clouds. 

The final panel in the bottom right of figure \ref{fig:cloudprop} shows the average H$_{2}$ ratio within clouds. 
Again using this with figure \ref{fig:cloud_h2} we can see that including feedback leads to clouds with higher ratios of molecular hydrogen, up to 70\%. This again suggested feedback is pushing clouds to higher densities promoting the formation of molecular hydrogen.
\subsection{Velocity dispersion of gas in our models compared to observations}
\label{section:larson}
\begin{figure}
    \centering
    \includegraphics[width=83mm]{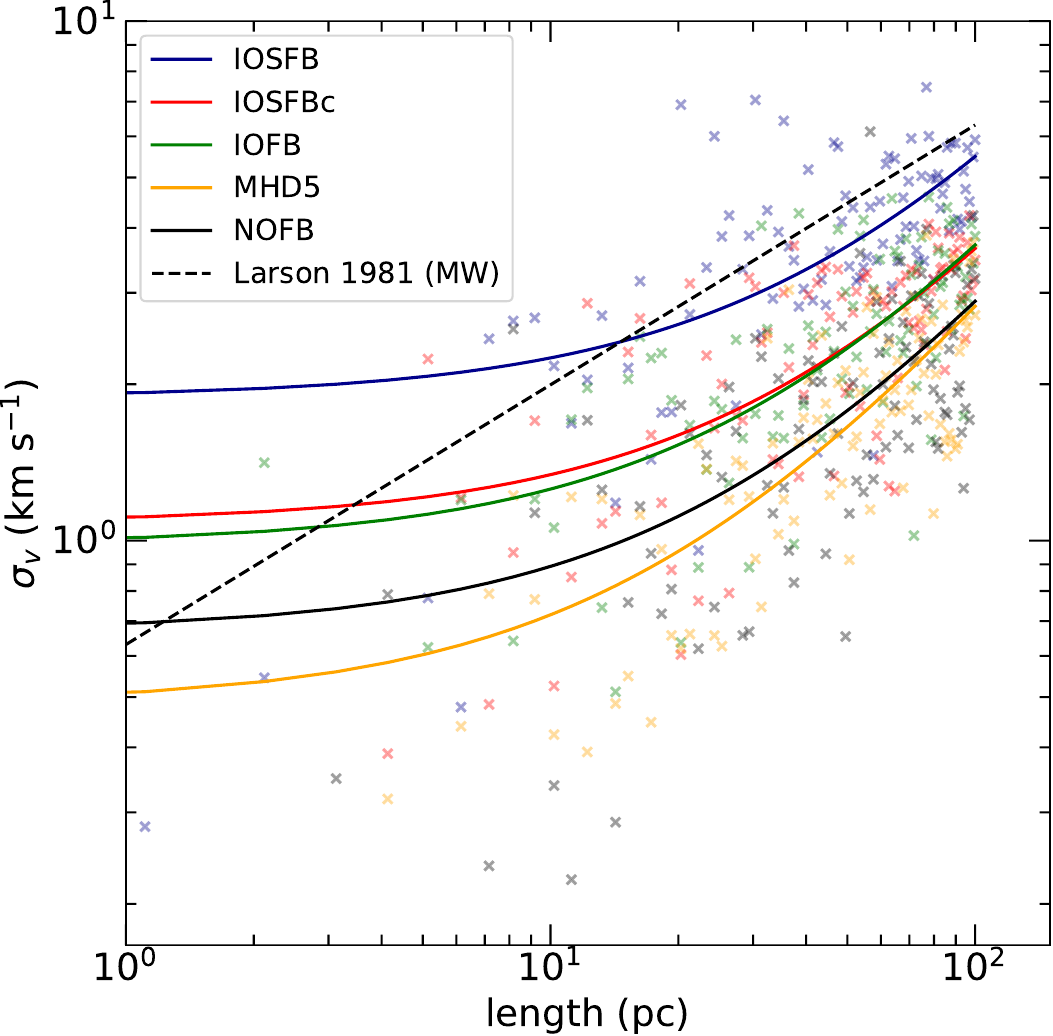}
    \caption{The Larson relation \citep{Larson1981} $\sigma_v = 10^{-0.2} l^{0.5}$ for velocity dispersion at a given scale, is plotted as the dashed line. This is compared to our Larson type relation for our no feedback and feedback models. The calculation for the Larson type relation is explained in section \ref{section:larson}. For each model the velocity dispersions are shown as points accompanied by a line of best fit.}
    \label{fig:larson}
\end{figure}
We compare our models to observed Milky Way cloud complexes with a Larson type relationship \citet{Larson1981} of the form $\sigma_v = 10^{0.2}l^{0.5}$ in a similar way as described in \citet{bending2022}. To reproduce this in our models we calculate the velocity dispersions for gas within randomly placed spheres of increasing radius from 1 to 100 pc. The velocity dispersion at a given scale are shown in figure \ref{fig:larson} and then a line of best fit is drawn for each model. It is clear that models IOSFB, IOSFBc and IOFB better fit the Larson type relation on the largest scales, meaning that feedback is important in driving gas motions at these scales. 
Including or excluding an initial population of stars (IOSFB and IOSFBc) affects the fitting to the Larson type relation. With an initial population of stars there are more supernovae occurring throughout the simulation. These supernovae drive turbulent motions at larger scales then the expansion of HII regions with ionising feedback only and so accounts for the increased dispersion in velocity at larger scales. Our model IOSFBc evolves in a similar way to IOFB since there are few supernovae throughout the simulation. This is reflected in the Larson type relation as they are similar on all scales. Without feedback the models NOFB and MHD5 do not fit the Larson type relation as well and are roughly similar on all scales as MHD does not affect the gas velocities significantly.
\section{Discussion and Conclusions}
\subsection{Impact of the initial region}
Our simulations show that, perhaps unsurprisingly, if the gas which is flowing into the region, is included over the timescales of interest, then the properties of the region accurately reflect those that it would have if it were still embedded in the whole galaxy. The simulations in \citet{Bending2020} likely (slightly) underestimate star formation at later times due to missing gas inflow. Choosing a region by tracing gas back (e.g. \citealt{dobbs2015re}) is one way to ensure gas flow is modelled, or the required locus of gas can be estimated analytically.


\subsection{Impact of the initial population of stars}
\label{section:impact_of_ini_pop}
\begin{figure}
    \centering
    \includegraphics[width=82mm]{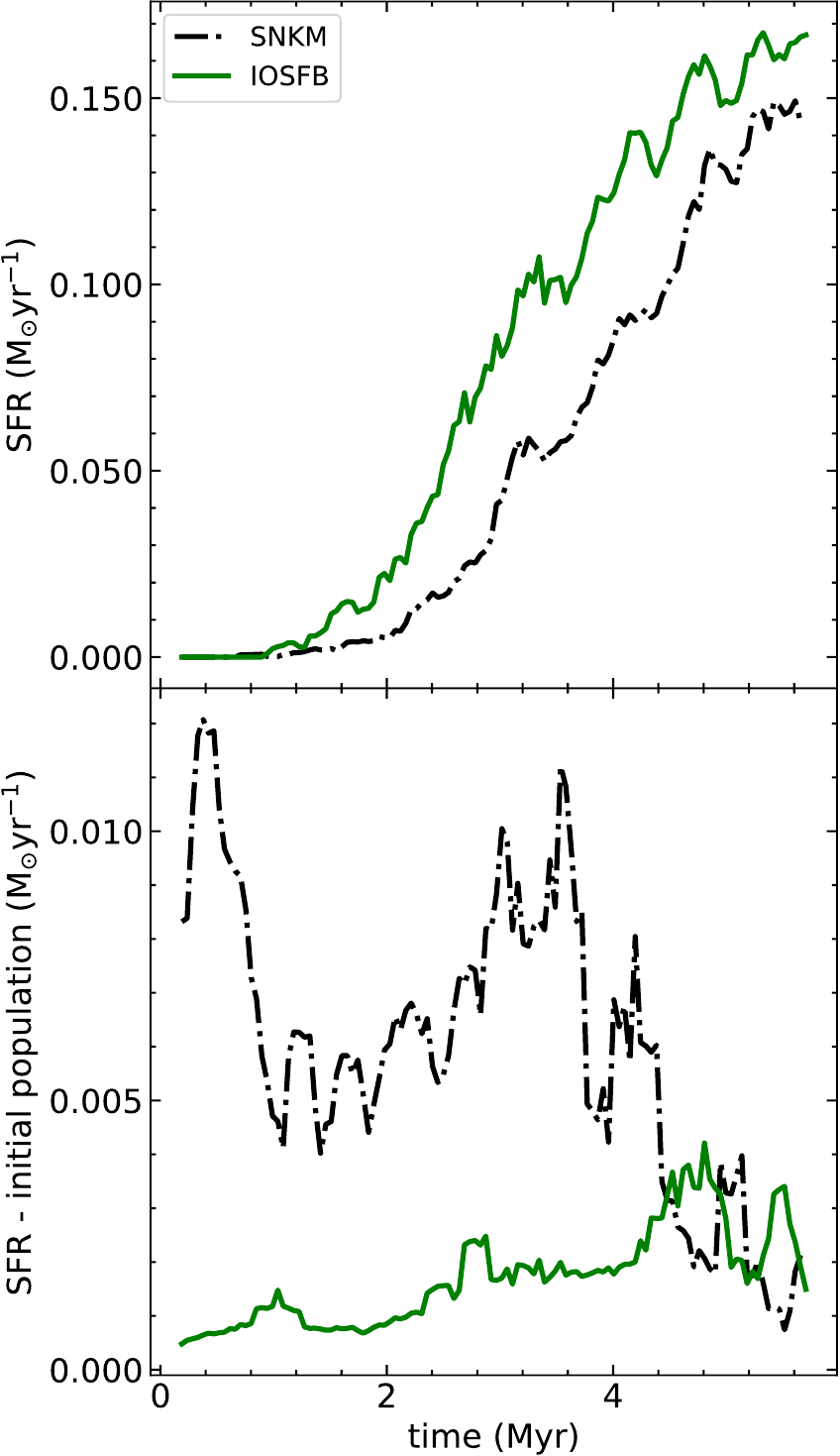}
    \caption{
    The star formation rate of IOSFB is shown in the top panel by the green line and the model SNKM is shown by the dot-dash black line. The bottom panel shows the SFR of the initial population, i.e. the accretion history of the sinks in both models.}
    \label{fig:snkm}
\end{figure}
An initial population of stars could potentially impact our simulations in two ways, firstly by producing warm ionised gas and reducing star formation, or by triggering the formation of new stars. Our simulations clearly find the latter. We find that for the first million years, the feedback has little effect. We expect the reasons for this are two fold, one being that HII regions need to grow and two the gas surrounding sinks is already relatively low density, due to feedback in the original galaxy simulation. Our initial population sinks do not accrete significantly throughout the simulation due to their low density environments, so their main effect is to trigger new star formation, through both ionisation and supernovae. We tested explicitly whether the distribution of the initial population has a significant impact by running a further model with the initial population preferentially placed in denser regions. To do so we take the same initial conditions as IOSFB (and same feedback during the evolution), but we redistribute the initial population into the denser regions. We locate dense gas by a cutoff of $3.39\times 10^{-22}$ gcm$^{-3}$ and randomly place sinks around gas particles above the cutoff within a maximum radius of 1 pc. This relocates sinks in spiral arm regions and dense filaments. We show the star formation rate of this model (named SNKM) compared with IOSFB in the top panel of Figure \ref{fig:snkm}. The star formation rates are very similar suggesting that the location of the initial population doesn’t make much difference. Looking at the bottom panel of Figure \ref{fig:snkm}, compared to IOSFB, there is more accretion at early times and less at later times onto the initial sinks, but the star formation associated with the initial population is minimal compared to the formation of new sinks, including those due to triggering. 

One aim of introducing an initial population was to potentially improve continuity between the galactic scale simulations and the resimulations. However if anything the initial population exacerbates this since the initial population increases triggered star formation, leading to greater disparity between the star formation rate in the global disc simulations and the local sub galactic simulations, and producing greater differences in structure. However the initial population of stars does drive motions on larger scales from the outset of the simulations, which otherwise is not necessarily captured.

\subsection{Magnetic fields}
Magnetic fields have a stronger impact on the rate of star formation than including an initial population of stars. Star formation is suppressed in our MHD model (MHD5) in the inter-arm and spiral arm regions. Small over densities in the inter-arm regions are smoothed out by the magnetic fields. 
This decreases the burst in star formation rate we see in the hydrodynamic feedback models after 5 million years by roughly a factor of two. Including MHD produces similar gas dynamics to our hydrodynamic model NOFB (no feedback).

\subsection{Summary of results}
\begin{itemize}
    \item[(i)] Testing our region extraction models by comparing SR1 and SR2 to the equivalent gas within extended regions SR1e and SR2e show surprisingly little difference in star formation rate. We note however morphological changes in the structure of region SR2, which we attribute to not capturing gas inflow to region SR2 at later times, due to poor choice of region (lack of gas perpendicular to the arm) and the smaller galactic radius.
    \item[(ii)] Including an initial population of stars unexpectedly increases the 
    star formation rate compared to having no initial population.
    Supernovae feedback from the initial population lead to increased star formation triggering. But the initial population themselves do not accrete very much throughout the simulation, since they reside in low density regions from prior feedback within the host galaxy model. This means, by accretion, their contribution to the rate of star formation is low.
    \item[(iii)] Supernovae from the initial population are important in driving gas motions on larger scales. The velocity dispersions on  scales > 100 pc  shows better agreement compared to the Larson relation \citep{Larson1981} than models without both forms of feedback. 
    \item[(iv)] Our MHD model shows that magnetic fields influence the rate of star formation more compared to an initial population of stars. Magnetic fields suppress star formation across the region due to extra magnetic pressure within star forming over densities. Here we only performed one comparison simulation with MHD, we plan to test different field strengths also with feedback in future work.
\end{itemize}


\section*{Acknowledgements} 
We thank the referee, for insightful feedback which has helped improve the paper. All authors in this work are funded by the European Research Council H2020-EU.1.1 under the ICYBOB project, grant number 818940. Our no feedback models where computed using the ISCA High Performance Computing Service located at the University of Exeter. The feedback models were performed with the DiRAC DIaL system, controlled by the University of Leicester IT Services, forming part of the STFC DiRAC HPC Facility (www.dirac.ac.uk). The equipment was funded by BEIS capital funding via STFC capital grants ST/K000373/1 and ST/R002363/1 and STFC DiRAC Operations grant ST/K001014/1. DiRAC is part of the National E-Infrastructure.
\section*{Data availability statement}
Upon reasonable request to the author, all data in this work will be shared.

\bibliographystyle{mnras}
\bibliography{refs} 






\bsp	
\label{lastpage}

\end{document}